\documentclass[aps,prb,reprint,superscriptaddress,showpacs,showkeys]{revtex4-1}

\usepackage{graphicx}
\begin{document}

\title{Apparent corrugation variations in moir\'e patterns of dislocated graphene on Highly Oriented Pyrolytic Graphite and the origin of the van Hove singularities of the moir\'e system}

\author{Dilek Y{\i}ld{\i}z}
\affiliation{\.{I}stanbul Technical University, Faculty of Sciences and Letters, Department of Physics, Maslak, 34469, Sar\.{i}yer, \.{I}stanbul, Turkey}
\author{H. \c{S}ener \c{S}en}
\affiliation{Bilkent University, Faculty of Science, Department of Physics, 06800, Bilkent, Ankara, Turkey}
\author{O\u{g}uz G\"ulseren}
\affiliation{Bilkent University, Faculty of Science, Department of Physics, 06800, Bilkent, Ankara, Turkey}
\author{O\u{g}uzhan G\"url\"u}
\email{gurlu@itu.edu.tr}
\affiliation{\.{I}stanbul Technical University, Faculty of Sciences and Letters, Department of Physics, Maslak, 34469, Sar\.{i}yer, \.{I}stanbul, Turkey}

\date{\today}

\begin{abstract}

Moir\'e patterns on Highly Oriented Pyrolytic Graphite surfaces due to dislocated graphene layers were studied. We observed that the apparent corrugations of the moir\'e patterns in scanning tunnelling microscopy images change as a function of tunnel junction bias. A simple geometric investigation of the atomic structure of the graphene layers generating moir\'e patterns revealed that different atomic arrangements due to different twist angles can result in similar geometric moir\'e periods such that, only a small fraction of the observed moir\'e periodicities may coincide with the real atomic periodicity generating the moir\'e system.  Ab initio calculations showed that the band structure of moir\'e patterns exhibit the fingerprints of those of bilayer graphene system preserving the Dirac cone. Moreover, our calculations in view of the correct atomistic modelling of the moir\'e patterns showed that van Hove singularities in twisted bilayer graphene system with varying angles have different origins in their respective band structures. Consequently, our results shed light on the graphene like behaviour of the top most dislocated graphene layer on HOPG surfaces by showing that the layer does not act alone but the final graphene bilayer has electronic properties partially resembling graphene. 

\end{abstract}



\maketitle

\section{Introduction}

Highly Oriented Pyrolytic Graphite (HOPG) can be considered as the drosophila of surface science, especially for the scanning probes. HOPG is a quite popular crystal in experimental studies especially in surface science because of its smooth and chemically inert surface. It is widely employed as a calibration sample for scanning tunnelling microscopy (STM) studies, as well as for educational purposes \cite{HOPG_student_Zhong2003, tomanek1, Moire1_Albrecht1988, HOPG_diffrent_atom_Wong_Durkan, cisternasHOPG_PrB}. HOPG is composed of Bernal stacked \cite{STM_Bernal_HOPG_Rong, HOPG_diffrent_atom_Wong_Durkan, cisternasHOPG_PrB} two dimensional hexagonal lattices, with carbon atoms at the lattice sites, popularly named as graphene layers \cite{layered_graphite_Kolmogorov2005, Tirlayer_HOPG_Graphene_rotation_Morell2013, turbostatic_graphenePanktarov2010}. Nearest neighbour carbon atoms in a layer are bonded strongly with each other, while the layers are coupled only by weak van der Waals interactions \cite{Novoselov_Type_2005, Geim2007Rise, layered_graphite_Kolmogorov2005}. 

Super periodic features were observed on HOPG surfaces \cite{Moire1_Albrecht1988, Moire2_Kuwabara, moire_3_Rong_1993, Durkan_Rev2005}. Due to the weak van der Waals bonding between the layers of HOPG, topmost layer may be shifted or rotated by mechanical or chemical means. As a result of the rotation of the top layer, super-periodic structures called as moir\'e patterns form on HOPG \cite{mechanical_moire_Rademan1998, Durkan_Rev2005, DurkanACNnano2009Mechanical, Wang_Wu2005_chem_surf_Sci}  surfaces.  Moir\'e patterns occur in nature and they are fascinating phenomena already optically, and when it comes to such structures at the atomic level, understanding their origin becomes even more tedious \cite{Durkan_Rev2005, deHeer2010trilayerSiC, Reasons_less_atom_1of3_Campanera}. Since the first observation of moir\'e patterns on HOPG surfaces with STM, their origin is still a matter of debate \cite{ Moire1_Albrecht1988, Moire_Nature_2011}. Despite the fact that these formations were investigated in numerous studies, they are rediscovered in graphene research \cite{CastroNeto_Bilayer_twist2007, MasslessFermion2007, Reasons_less_atom_1of3_Campanera, cisternasVargas2008, Conrad_WdH_2008_SİC_grapphene_like, VanHove_HOPG_moire_Nature_Phy, STS_Graphene_HOPG_PRL2009, Luican2011_LLS, Moire_Nature_2011, Moire_metal_scirep}. Due to its inevitable relevance to graphene based devices the importance of dislocated graphene on other graphene layers on various systems drew serious attention \cite{vonKlitzingSiC_QHE2011, Conrad_WdH_2008_SİC_grapphene_like, Novoselov2004Science, Nanolett2010_magaud, SiC_device_1}. Still the answers to the questions like ''When does the top most layer behave like graphene?'', or ''Will that layer have other electronic properties'' remain yet to be settled. Localized effects also have to be understood and utilized or eliminated for any device application.

By means of Angle Resolved Photoemission Spectroscopy (ARPES) \cite{ARPES_no_VhV, SiC_Arpes2}, Landau Level Spectroscopy (LLS) \cite{Luican2011_LLS, VanHove_HOPG_moire_Nature_Phy, STS_Graphene_HOPG_PRL2009} and Quantum Hall Effect (QHE) \cite{vonKlitzingSiC_QHE2011, QHE_thoery_TBG_Koshino2012} measurements the graphene like behaviour of top most layers were studied. The electronic structure of the moir\'e patterns and the van Hove (VHV) singularities observed on these systems were also intensively studied both theoretically and experimentally \cite{VanHove_HOPG_moire_Nature_Phy, Brihuega2012PRL, Luican2011_LLS, MoireBands_PNAS2011, Landgraf2013, vHv_dirac_Chu2013, cisternas3_2012, QHE_thoery_TBG_Koshino2012, Yan2014}. ARPES results suggest that the top most layers of Graphene/SiC system always behave like graphene \cite{ARPES_no_VhV} while the observed VHV singularities and LLS measurements suggest otherwise, by simply showing a rotational dependence \cite{Nanolett2010_magaud, Brihuega2012PRL, Luican2011_LLS, Yan2014}. Moreover, the QHE measurements support LLS and VHV measurements. Although moir\'e patterns and their geometric structure were considered in all studies, effect of the real atomic conformation of the moir\'e patterns was not discussed in detail. The reason of the VHV singularities from the band structures of the moir\'e patterns is not complete yet, either. The lack of any moir\'e structure observation on HOPG by atomic force microscopy (AFM) is yet to be understood. So the question remains: ''Are these structures comparable to the optical moir\'e patterns or are they an outcome of the electronic structure of these systems?'' The variation of the apparent corrugations of the moir\'e patterns on HOPG surfaces observed in STM measurements as a function of tunnel junction parameters were not investigated.  In almost all the studies the corrugation was reported to be constant as a function of tunnel junction bias \cite{Durkan_Rev2005} but for one \cite{Wong2011}, where the change of moire corrugation as a function of distance from defects, in resamblance to Friedel oscillations, was discussed. Inexistance of the corrugation variation as a function of bias voltage was reported for moir\'e patterns observed on SiC \cite{Brihuega2012PRL} as well. 

Besides the experimental observations on HOPG and Graphene/HOPG systems, understanding these data became a challenging issue for theoretical studies. For instance, while the inspection of the HOPG surface became a benchmark for STM studies \cite{tosatti_HOPG_STM}, explaining the surprising triangular structures in spite of the hexagonal lattice of graphite, which has two atoms in the surface unit cell, became a formidable challenge. Nevertheless, this was successfully answered by realizing that the \textbf{b} site carbon atom which sits at the center of underlying hexagonal carbon ring of the second layer has slightly higher local density of states (LDOS) around Fermi energy compared to the carbon atom sitting at the \textbf{a} site (see Figure~\ref{fig:figur1}a) of  AB stacked graphite \cite{tomanek1, tomanek2, cisternasHOPG_PrB, PRL_Carbon_imaging2011}. Consequently, the first attempts to describe the reported STM images of moir\'e superstructures were based on this fact \cite{moire_3xhie}, which had foreseen completely opposite bright regions. This was corrected by the model developed from the ab initio calculation for the DOS of AA, AB, and ABC stacked graphites \cite{moire_3_Rong_1993}. This successful latter model was extended to investigate in detail the moir\'e patterns on graphite \cite{Reasons_less_atom_1of3_Campanera}. Obviously, all these approaches rely on perfect stacking results without performing a full Density Functional Theory (DFT) calculation on the rotated superstructures. Later on, some calculated STM images using different DFT based methods were reported on moir\'e superstructures due to twisted graphene layers on HOPG surfaces \cite{cisternasVargas2008, cisternasHOPG_PrB, cisternas3_2012}.

In this paper we report on the atomic structure of the moir\'e patterns on HOPG surfaces and we unequivocally show that many of the similar appearing moir\'e patterns in STM images do not have the same atomic sub-structure. We report the apparent height corrugation variations of moir\'e patterns as a function of tunnel junction bias measured with STM on HOPG surfaces for the first time. Our STS measurements and DOS calculations show that moir\'e structures behave more metallic than HOPG, but not graphene like. Our results show that the information on the real atomic structure leading to each moir\'e pattern is clearly invaluable for understanding the electronic structure of this system, as they show for the first time that van Hove singularities observed on different moir\'e systems do not have the same origin in their respective band structure. 

\section{Experimental: Preparation and measurement of moir\'e patterns on HOPG}

Various methods have been proposed for the preparation of moir\'e patterns on HOPG surfaces \cite{mechanical_moire_Rademan1998, Durkan_Rev2005, DurkanACNnano2009Mechanical, Wang_Wu2005_chem_surf_Sci} besides the random encounters on HOPG as well as on other graphitic systems like SiC \cite{Brihuega2012PRL} or multilayer graphene grown by Chemical Vapor Deposition (CVD) and transferred on to other substrates \cite{Luican2011_LLS}. Reported chemical methods require time and extended processes, by means of which the HOPG crystal may be damaged permanently. We have prepared HOPG surfaces with moir\'e patterns on them by means of a very simple and repeatable method. We have discovered that simply by drop casting cyclohexane on to HOPG surfaces (SPI$\textsuperscript{TM}$ grade 2 and 3) and letting the samples dry in a clean environment, many different moir\'e structures can be obtained \footnote{Understanding the formation of moir\'e patterns on HOPG surface due to cyclohexane drop casting and drying process is beyond the scope of this present work and will be discussed elsewhere}. Pure and fresh (VWR$\textsuperscript{TM}$ cyclohexane, HPLC grade) is used for each sample preparation. HOPG samples were thoroughly investigated by STM prior to cyclohexane drop casting. After the drop cast and drying process, we have checked all the surfaces by means of micro Raman spectroscopy for probable contamination or residue from the cyclohexane drop casting and drying method. There were no observable remnants. Rarely (1 in 50 tunnelling zones) native moir\'e patterns were observed on both pristine SPI grade-2 and grade-3 HOPG crystals. All the measurements reported here were performed under ambient conditions by use of a Nanosurf Easyscan2$\textsuperscript{TM}$ STM and mechanically cut $Pt_{80}$$Ir_{20}$ tips. The experimental conditions were carefully controlled and a consistency checklist was developed \footnote{Sample preparation procedure and the procedure we have developed to obtain consistent data is further explained in the supporting information}. Many different moir\'e samples were prepared and investigated by use of different tips.  On the same sample reliable and consistent measurements for longer than half a year were possible.

Since the site \textbf{a} atoms of the HOPG surface have less charge density compared to site \textbf{b} atoms (Figure~\ref{fig:figur1}a) STM gives triangular-atomic resolution on HOPG surfaces \cite{cisternasHOPG_PrB}, where only the \textbf{b} atoms were observed. In some cases by use of quite high tunnelling current measurements, \textbf{a} atoms can also be observed by low temperature STM \cite{tip_effect_stm_res2012}. Intercalation of chemicals may result in hexagonal resolution \cite{Wang_Wu2005_chem_surf_Sci} on HOPG surfaces as well. On the moir\'e patterned graphene/HOPG samples we always obtained triangular resolution on pristine HOPG terraces just next to the moir\'e zones. More importantly, as shown in Figure~\ref{fig:figur1}c and d, it was possible to achieve atomic resolution on moir\'e patterns, simultaneously with the moir\'e structure observation. This was possible within a broad bias range (from 50 mV up to 800 meV). Still it was only possible to observe triangular atomic resolution on pristine HOPG terrace just next to the moir\'e zone, only by 50 meV, with the same tip during the same measurement. On clean monolayer graphene, transferred on to dielectric substrates, it was possible to obtain hexagonal-atomic resolution in STM measurements \cite{AtStructTriHexNanoLet, SuperLattDiracSTMHeX}. The lack of total hexagonal resolution, but rather a partial hexagonal resolution on moir\'e patterns indicate that top layer does not behave like graphene (Figure~\ref{fig:figur1}c and d). Rather it must be the property of the top most layers resulting in the moir\'e patterns. 
\section{Theoretical}
\subsection{Methods}
\begin{figure}
\includegraphics[width=80mm]{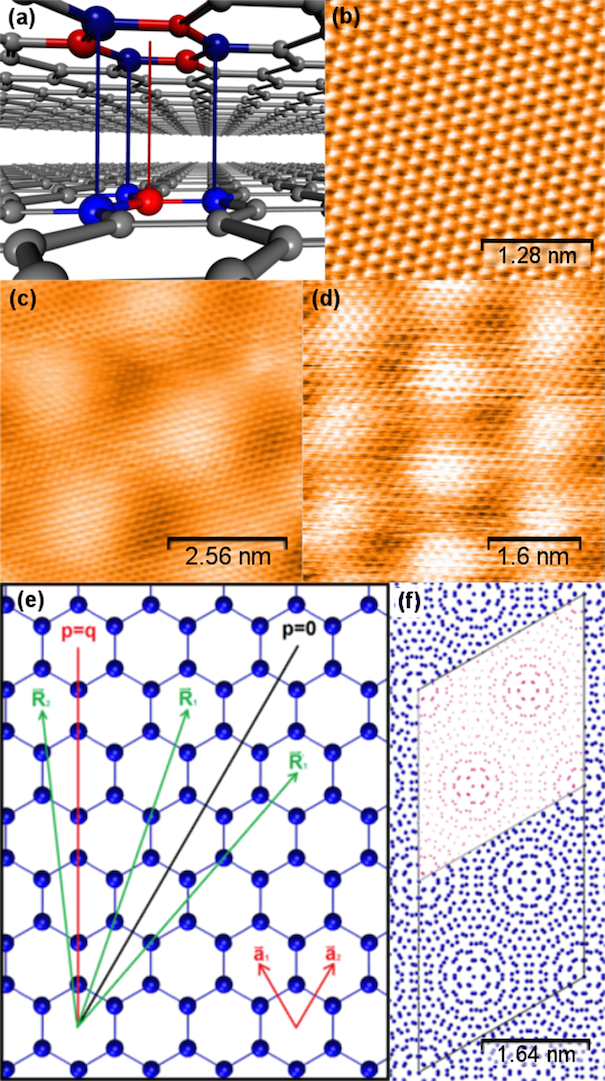}
 \caption{\label{fig:figur1}(a) HOPG model; \textbf{a} type atoms are shown in blue and \textbf{b} atoms are indicated with red. (b) Atomic resolution STM image on HOPG, only \textbf{b} atoms are visible by STM ($I_{t}$= 0.5 nA, $V_{b}$= 50 mV (sample bias)). (c) and (d) are STM images of different moir\'e patterns. Mark the atomic resolution on them. (c)  $I_{t}$= 0.45 nA, $V_{b}$= -200 mV(sample), observed GMP=2.54 nm   (d) $I_{t}$=0.4 nA, $V_{b}$=-50 mV (sample), GMP=1.6 nm. (e) Generation of the moir\'e pattern (see text for details). (f) Generated moir\'e pattern $p_{1}$=1 and $q_{1}$=11 (Rotation angle=8.61$^{\circ}$, GMP=1.64 nm and RMP=2.84 nm). Inset image shows calculated charge densities for this pattern. Observe the difference in the centers of super periodic formations.}
\end{figure}
Efforts for understanding such intricate nature of the moir\'e structures by electronic structure calculations of the twisted graphene layers for different rotation angles were performed using different methods. The major difficulty of computational studies was the number of atoms within the unit cell, which depends on the rotation angle and increases enormously with decreasing angle \cite{ Reasons_less_atom_1of3_Campanera}. Because of this, several studies were based on continuum theory, which exploited the effective Hamiltonian describing the linear dispersion of Dirac cones of low energy excitations \cite{CastroNeto_Bilayer_twist2007, CastroNeto2, mele2010, falco}. There were tight binding calculations for the rotations, which can be described by a few thousands atoms within the unit cell \cite{Nanolett2010_magaud, magaud2, Landgraf2013}. Accurate but computationally expensive ab initio calculations were very limited and only reported for relatively large rotation angles \cite{Nanolett2010_magaud, magaud2, jung}. Three different regimes emerged from these studies \cite{Nanolett2010_magaud, magaud2}: (1) For large rotation angles, two layers decouple from each other, so the linearly dispersed bands of Dirac cones of top and bottom layer remain unchanged \cite{MasslessFermion2007, pankratov1}. (2) In the intermediate regime, for rotation angles between 1$^{\circ}$ and 15$^{\circ}$, Fermi velocity renormalizes since the slope of linear Dirac bands decreases with decreasing rotation angle. Furthermore, the Dirac cone of the top layer and the rotated cone of the second layer intersect at finite energy so van Hove singularities emerge from these ''hybridized'' bands and clearly they depend on the rotation angle \cite{CastroNeto_Bilayer_twist2007, CastroNeto2, Nanolett2010_magaud, magaud2}. (3) Flat bands and vanishing velocities were reported for magical angles for rotations smaller than 1$^{\circ}$. \cite{Nanolett2010_magaud, magaud2, MoireBands_PNAS2011}.

We used VASP for the first-principle plane-wave calculations that employ density functional theory by the projected-augmented-wave potentials \cite{kresse1, kresse2}. GGA is used to express the exchange-correlation potential and plane-wave cut-off energy is set to 500 eV. The Brilluoin zone integrations were carried out by using k-point meshes based on Monkhorst-Pack scheme defined according to size of the system (moir\'e) unit cell, and the smearing parameter was set as 0.08 eV. The energy convergence was within $10^{-5}$ eV accuracy while relaxing the systems electronically. In view of our calculations here we discuss the properties of moir\'e superstructures from the band structure and total density of states (DOS) as well as the STM images calculated by integrating the local density of states from the bias voltage ($V_{b}$) to the Fermi energy ($E_{f}$) as suggested by Tersoff and Hamann \cite{tersoff-1, tersoff-2}. Note that, it is necessary to use extremely dense k-point meshes in DOS calculations in order to avoid artificial oscillations and clearly show the features of van Hove singularities emerging in the DOS of the moir\'e systems. In order to understand the effects of bottom layers on the charge density profile of the top layer we performed some pre-calculations on AB stacked graphite. We increased the number of layers from two to six gradually and observed no significant difference in the charge density profile of the top layer around the Fermi level. Therefore, we describe moir\'e patterns by twisting a bilayer graphene system. We also performed tight binding (TB) calculations using the well-tested coupling parameters \cite{tomanek2} in order to point out the critical differences with respect to ab initio calculations.

\subsection{Modelling the moir\'e patterns: Real Moir\'e Periodicity vs. Geometric Moir\'e Periodicity}

\begin{figure}
\includegraphics[width=85mm]{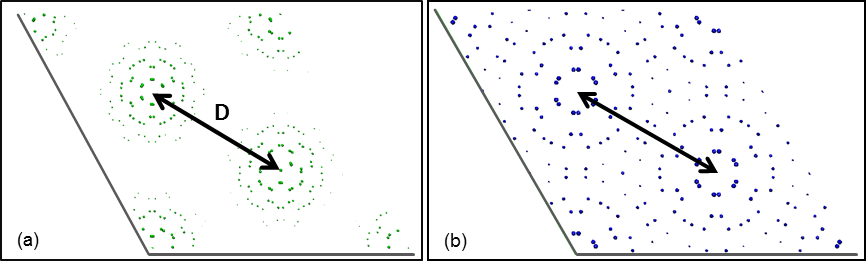}
  \caption{\label{fig:figur22} (a) Calculated charge density for  $\theta_{rotation}$=8.61$^{\circ}$, GMP = 1.64 nm and RMP = 2.84 nm. Here D is the GMP but RMP is  $\sqrt3$D. (b) Charge density plot of a case where RMP=GMP=10.7 nm and $\theta_{rotation}$ =13.17$^{\circ}$. In this case the moir\'e unit cell is much smaller and consequently the systems has far less atoms to consider in ab initio calculations.}
\end{figure}

We apply a practical scheme to describe the moir\'e superstructure unit cell of twisted bilayer graphene as follows: Consider AA rather than AB stacked bilayer graphene as shown in Figure~\ref{fig:figur1}e where only the top layer atoms of the model are visible. We can write any lattice translation vector of the top layer as    $\overrightarrow{R_{1}}= p_{1} \overrightarrow{ a_{1}}+q_{1} \overrightarrow{a_{2}} $   by using the lattice vectors  $\overrightarrow{ a_{1} }$ and $\overrightarrow{ a_{2} }$ , where ${p_{1}}$ and ${q_{1}}$ are integers among themselves. In order to generate the unit cell of twisted bilayer graphene, we need to find the corresponding lattice translation vector   $\overrightarrow{ R'_{1}} $ of the bottom layer.   $\overrightarrow{ R'_{1}} $  can be defined as the mirror symmetry of  $\overrightarrow{ R_{1}} $ with respect to p=0 line. Hence, $\overrightarrow{ R'_{1}} $  becomes $-p_{1} \overrightarrow{ a_{1} } + ( q_{1} + p_{1}  ) \overrightarrow{ a_{2} }$ . The angle between the p=0 line and $\overrightarrow{ R_{1}} $ and the angle between the p=0 line and $\overrightarrow{ R'_{1}}$ are same. Therefore, these two vectors coincide with each other on p=0 line by rotating the two layers towards this line by $ \theta$. (The angle of rotation of the top layer with respect to the lower layer within this scheme can be defined in terms of $p_{1}$ and $q_{1}$ as $ \theta_{rotation} = 2 \theta = 2 arctan ( \sqrt3 p_{1} / 2q_{1}) $.) This rotated vector points to a lattice point of the twisted bilayer. The other lattice vector of this two dimensional lattice of twisted bilayer graphene can be calculated by rotating this lattice vector by $\pi$/3 or 2$\pi$/3. These two vectors define the primitive unit cell of moir\'e superstructure unless $q_{1}$-$p_{1}$=3k, k being an integer. If that is the case, the primitive unit cell can be defined from a shorter lattice vector as  $\overrightarrow{ R_{2} } =((2 p_{1}+ q_{1}) \overrightarrow{ a_{1}}/3)+(( q_{1}-p_{1}) \overrightarrow{a_{2}}/3)$   of which length is $|\overrightarrow{R_{1}}|/ \sqrt3$. Consequently, when ${p_{1}}$ is equal to one, 1/3 of all cases are special where real and observed (geometric) moir\'e periodicities (RMPs and GMPs) are equal, while for the rest real moir\'e periodicity is  $\sqrt3$ times larger than the observed super-periodic structure (see Figure~\ref{fig:figur1}f and Figure~\ref{fig:figur22}). If $p_{1}$ is not equal to one, then there are even less cases where GMPs are the same as RMPs. We have performed first principles calculations for $p_{1}=1$ and $q_{1}=5, 7, 10,11,12,16$ cases.

In Figure~\ref{fig:figur1}f the rotation angle of the graphene layer is 8.61$^{\circ}$ ($p_{1}$=1 and $q_{1}$=11). Upon investigating the centres of the round formations in such figures, they are observed to differ among themselves. Here the definitions of Real Moir\'e Periodicity (RMP) and Geometric Moir\'e Periodicity (GMP) become clear. In Figure~\ref{fig:figur22}a and b the calculated charge densities for two different systems, one being an RMP is different than GMP case and the other being a GMP is the same with RMP case, are given. In STM images such minute differences are almost impossible to identify and even in the model this small detail is easy to miss. In the literature moir\'e periodicities are assumed to follow the relation: D=a/(2sin($\theta$/2)) (where a is the lattice parameter) for bilayer graphene system. However, this relation only indicates the distance (D) between the two bright spots observed, which is the GMP (see Figure~\ref{fig:figur22}). Consequently, it would not be possible to deduct the real rotation angle using only the moir\'e periods resolved in STM images. In such a case one cannot expect similar electronic structure for two dissimilar atomic configurations. Still the DOS of these systems show some surprising similarities, like a dependence of the energetic positions of van Hove (VHV) singularities on the rotation angle, which is a really lucky coincidence, which will be discussed below.

We present two examples on the differences and similarities between RMPs and GMPs: When $p_{1}$=1 and $q_{1}$=7 and $p_{1}$=2 and $q_{1}$=15 cases are considered, their GMPs are 10.72 \AA  and 11.4 \AA  respectively, which is pretty close and almost impossible to identify from STM images. However, their RMPs are 10.72 \AA  and 39.59 \AA  respectively, which shows the drastic difference between the amounts of atoms to be considered in the unit cells of both systems. When, the rotation angles for these cases are compared, the required angle for $p_{1}$=1 and $q_{1}$=7 case is 13.17$^{\circ}$, whereas this is 12.35$^{\circ}$ for $p_{1}$=2 and $q_{1}$=15. Due to such minute differences in angles, and since in an STM data only GMPs can be clearly identified, the real value of the rotation angle may not be evaluated from STM data even if atomic resolution was achieved.

\section{Results and Discussions}
\subsection{Corrugation variation on moir\'e patterns as a function of tunnel junction bias}
\begin{figure}
\includegraphics[width=85mm]{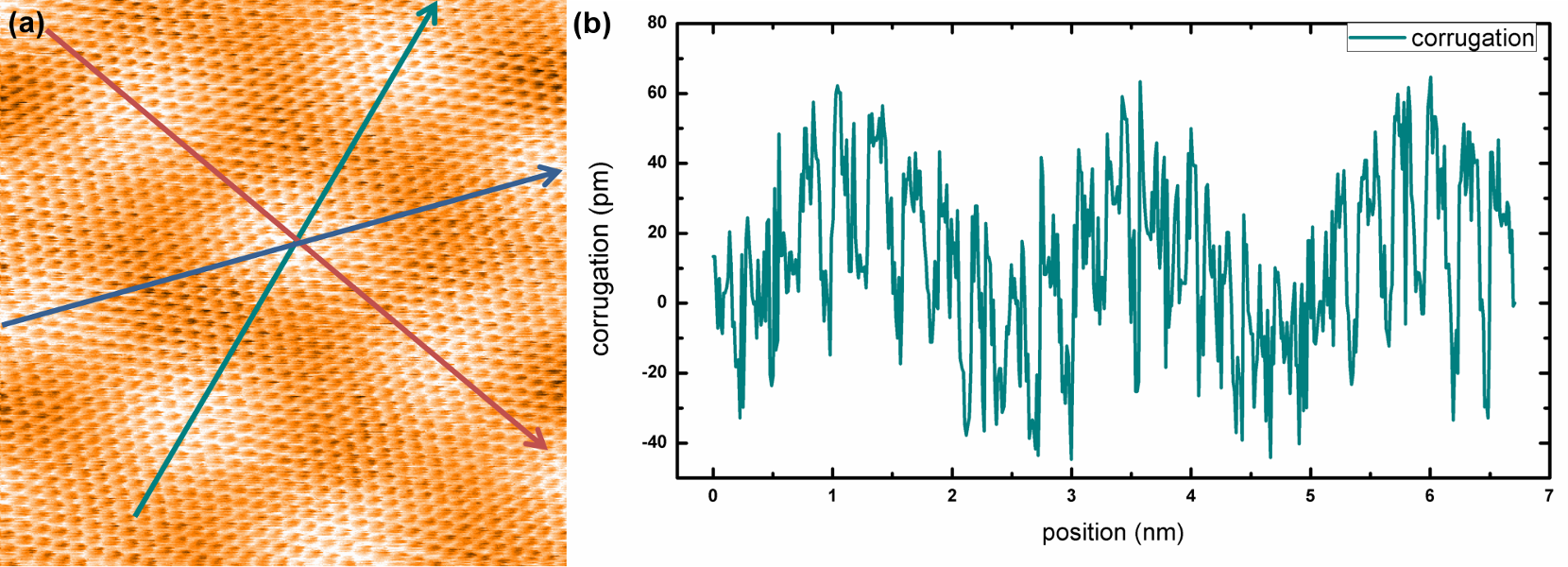}
  \caption{\label{fig:figur33} 6.4 nm  $\times$ 6.4 nm area scanned with $V_{b}$= -45 mV (sample bias), $I_{t}$= 0.5 nA. Here the line scan on the moir\'e pattern shows the atomic corrugation along with the moir\'e corrugation}
\end{figure}
\begin{figure*}[ht] 
\includegraphics[width=6in]{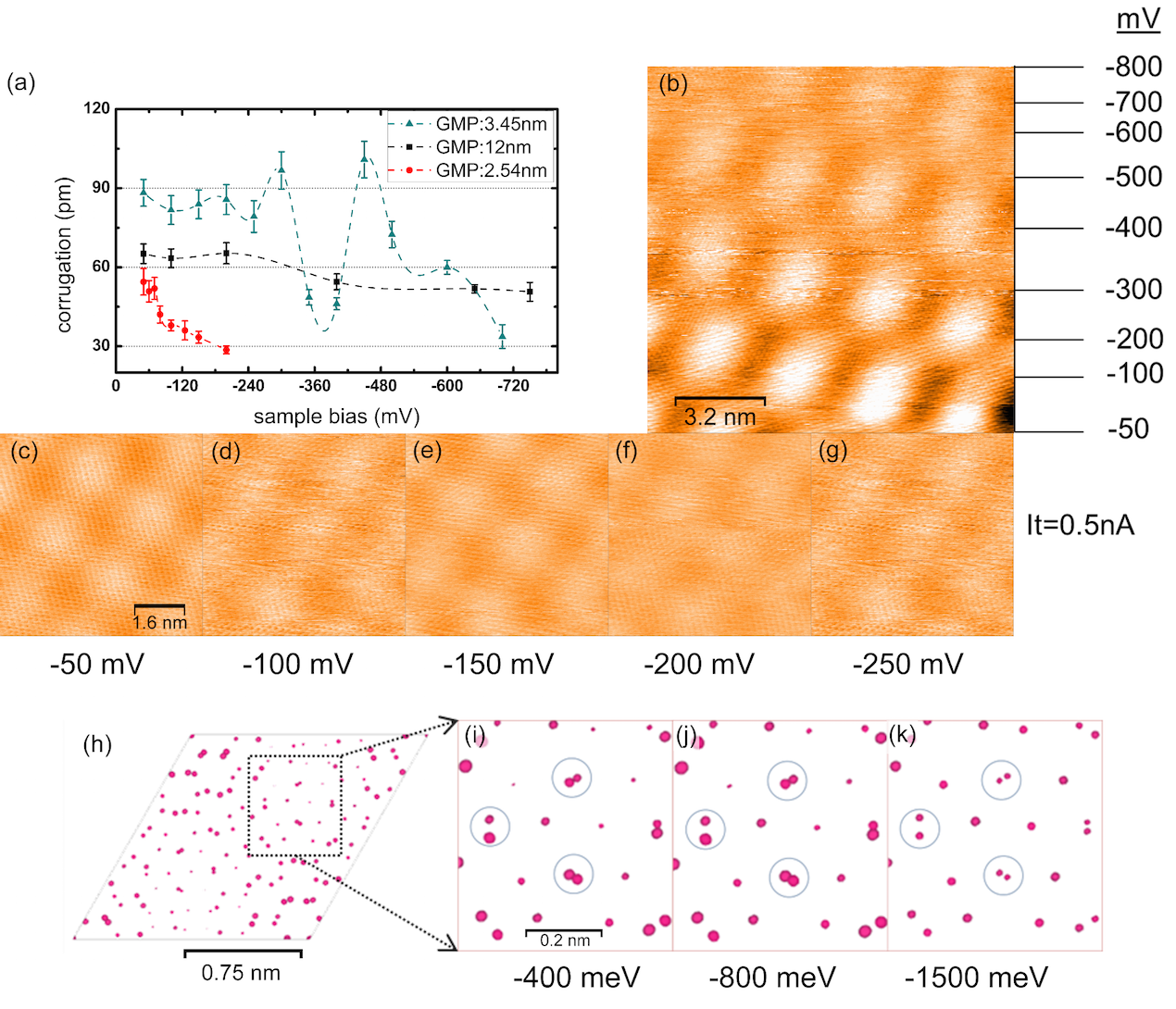}
  \caption{\label{fig:figur2}(a) Bias dependent z-corrugation change of three different moir\'e patterns, of  GMPs 3.45 nm, 12 nm and 2.54  nm. Apparent  z-corrugation of GMP=2.54 nm decreases with increasing bias while that of moir\'e with 3.45 nm GMP does not exhibit a monotonic decrease and GMP=12 nm tends to stay almost constant in the same voltage range. Tunneling currents used in the measurements  of 3.45 nm moir\'e is 0.5 nA, 12 nm is 0.7 nA, and 2.54 nm moir\'e is 0.5 nA. (b) During one of the scans on GMP=3.45 nm moir\'e structure, bias was changed at several lines and moir\'e corrugation  was observed to decrease with increasing bias (It=0.5 nA, $V_{b}$ was changed from -50 mV to -800 mV during scanning). (c)  $V_{b}$=-50 mV, (d) $V_{b}$=-100 mV, (e) $V_{b}$=-150 mV, (f) $V_{b}$=-200 mV , (g) $V_{b}$=-250 mV ($I_{t}$=0.5 nA for (c), (d), (e), (f), (g)). (h) Charge density contour plots of $p_{1}$=1 $q_{1}$=10 at -400 meV. (i) zoom on to a small area - 400 meV (j) -800 meV (k) -1500 meV. Follow the charge density contours in the circled areas to observe the apparent size changes.}
\end{figure*}
\begin{figure*}[ht]
 \center
\includegraphics[width=7in]{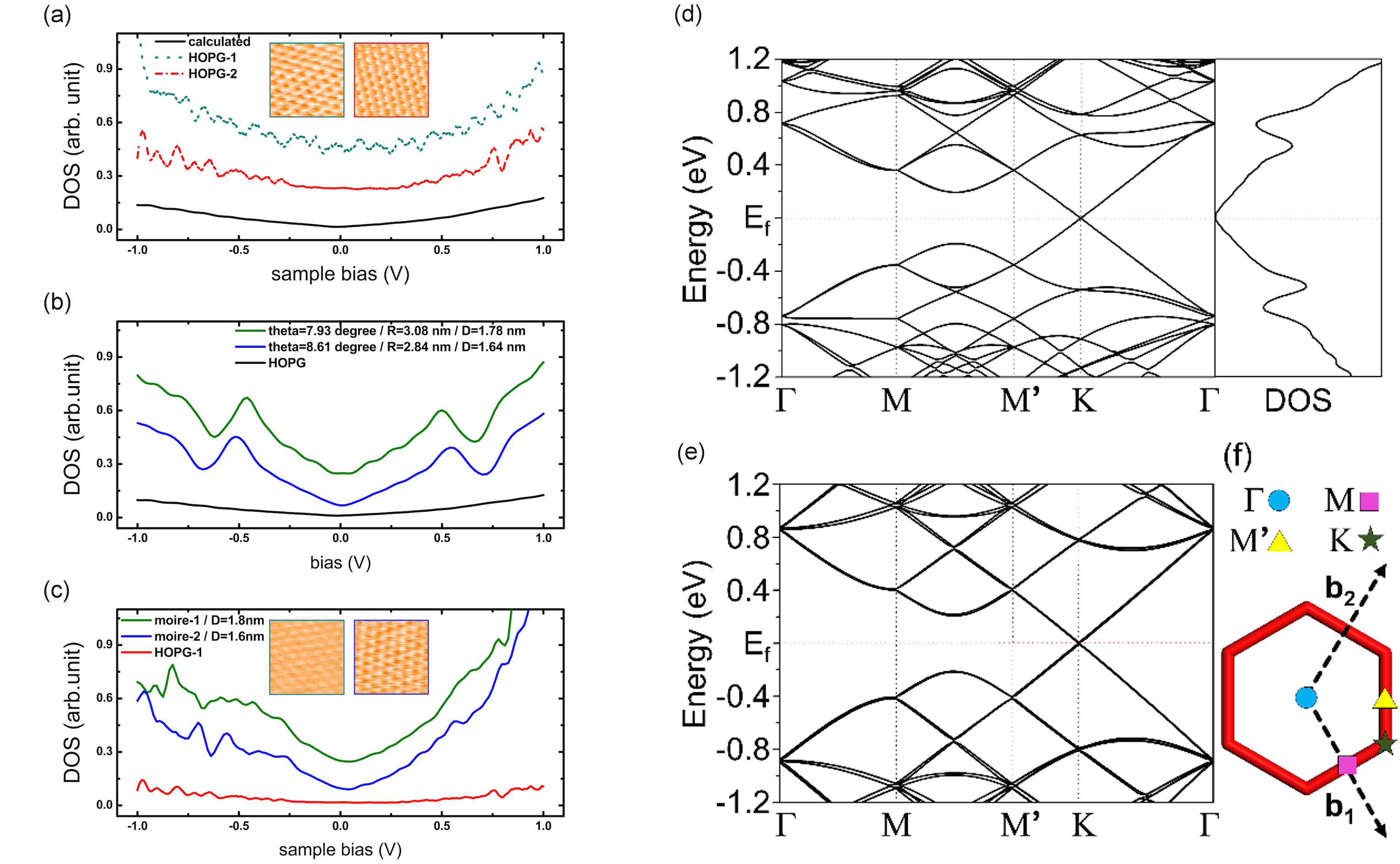}
  \caption{\label{fig:figur3}(a) Calculated and measured DOS plots of HOPG. Size of the insets are 1.6 nm $\times$ 1.6 nm. $I_{t}$=0.5 nA and $V_{b}$= -50 mV (sample bias) (b) Calculated DOS plots of HOPG, $p_{1}$=1 $q_{1}$=11 moir\'e and $p_{1}$=1 $q_{1}$=12 moir\'e. (c) Measured DOS of two moir\'e samples and HOPG. GMP of the first moir\'e is 1.8 nm and the second moir\'e is 1.6 nm. The size of the inset images are 12.8 nm $\times$ 12.8 nm. Tunneling parameters; $I_{t}$=0.4 nA and $V_{b}$=-50 mV . Mark that moir\'e zones behave metallic compared to the HOPG surface. (d) Calculated band structure and density of states of $p_{1}$=1 $q_{1}$=11 moir\'e structure with ab initio planewave pseudo potential method. (e) Calculated band structure of $p_{1}$=1 $q_{1}$=11 moir\'e structure with tight binding method. (f) High symmetry points used for the calculation of band structures inside the moir\'e Brillouin zone. High symmetry points are marked as a guide to the eye.}
\end{figure*}
We collected STM images on various moir\'e patterns with varying tunnel junction biases while the tunnelling current was kept constant. During the measurements moir\'e structures were observed to be structurally stable. It was possible to observe atomic resolution on the moir\'e patterns and investigate both atomic corrugation along with the moir\'e corrugation by line scans as presented in Figure~\ref{fig:figur33}. For each moir\'e zone many STM images were obtained at different bias values while keeping the tunnelling current constant. Than on every image several line scans were taken along 3 symmetry directions. From these data corrugation values were obtained. This was performed at least on ten different positions on the same image and on all the moir\'e structures at every bias voltage measured. Various tunnel junction biases were used from 50 mV to 800 mV without damaging the moir\'e structures or the STM tip. In Figure~\ref{fig:figur2}a we show the variation of the apparent moir\'e corrugations observed on three different moir\'e patterns (with different GMPs) as a function of the tunnel junction bias. In Figure~\ref{fig:figur2}c-g we show that the atomic resolution was observable in a broad range of biases on this moir\'e pattern. Decreasing corrugation with increasing absolute value of tunnel junction bias was observed for these three super-periodic structures. Measured GMPs of these structures were 2.54 nm, 3.45 nm and 12 nm. While observed corrugation of two samples decreased with increasing tunnel junction bias, the third sample did not only show a decrease but also a varying behaviour in its apparent corrugation (both increasing and decreasing) as a function of increasing bias. In order to investigate the physical meaning of this apparent corrugation variation we have made a comparative charge density plot of the ${p_{1}}$=1 ${q_{1}}$=10 moir\'e system with RMP=GMP=1.5 nm, as a function of different energy windows from $E_{f}$. These can be compared to different tip biases in STM measurements. Upon careful investigation of the local charge density distribution on the surface for the given energy window (i.e. Tersof-Haman calculated STM images), it is apparent that relative size of some of the charge density contours increase and than decrease with increasing bias  (Figure~\ref{fig:figur2}i,j and k). It is clear that STM tip cannot attain such a high lateral resolution but rather an averaged out image was obtained. Still, such variations in the electronic structure can be observed as corrugation variations in STM images. Both our experimental and theoretical data indicate that the apparent corrugation of the super periodic structures (i.e. moir\'e patterns) varies with varying bias and this variance does not necessarily follow a decay. Such STM data and ab initio calculations show that the top most layer does not behave like graphene but the top two layers that generate the moir\'e patterns on the HOPG surface have a new electronic structure and the resulting surface may partially resemble graphene.

\subsection{Density of states of the moir\'e patterns}

\begin{figure*}[ht]
 \center
\includegraphics[width=6in]{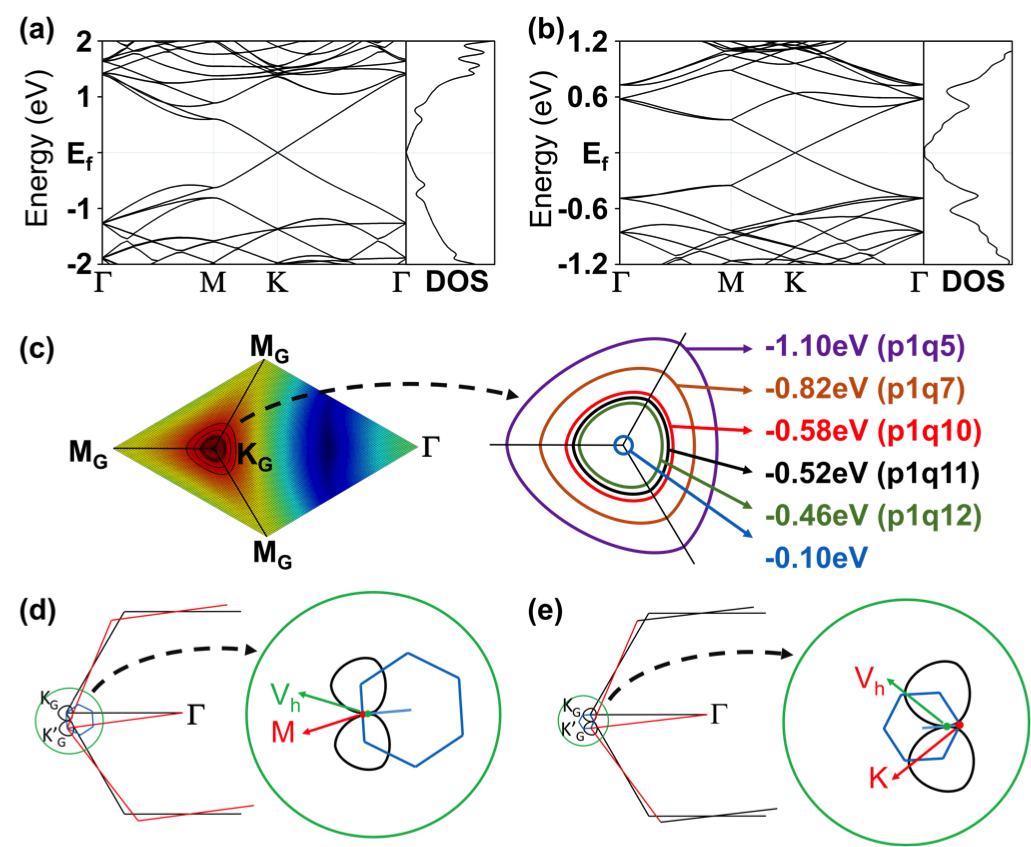}
 \caption{\label{fig:figur4}Calculated band structure (ab initio planewave pseudopotential method) and DOS of  (a) $p_{1}$=1 and $q_{1}$=10 moir\'e structure, (9.43$^{\circ}$, special case of 1/3, where RMP=GMP=14.96 \AA) has VHV at about 0.58 eV. (b) $p_{1}$=1 and $q_{1}$=12 moir\'e structure, (7.93$^{\circ}$, GMP=17.8  \AA and RMP=30.82  \AA) has VHV at about 0.46 eV. (c) Energy band diagram of $\pi$ band of graphene within the irreducible BZ. Energies corresponding to the VHV singularities of various moir\'e structures are indicated. (d) The relation of bilayer graphene and moir\'e BZs at the emergence of Van Hove singularity of $p_{1}$=1 $q_{1}$=10, and (e) of $p_{1}$=1 $q_{1}$=11 moir\'e structures. (see the supplementary video for further information.)}
\end{figure*}

We performed scanning tunnelling spectroscopy (STS) measurements on moir\'e patterns as well as on clean HOPG terraces just next to the moir\'e patterned zones (Figure~\ref{fig:figur3}). The measured I-V curves were used for numerically calculating dln(I)/dln(V) vs. V curves which give the DOS of the surface of interest as commonly performed in the literature \cite{gurluPRLGe-sts}. The immediate observation in the STS data is the relative metallic behaviour of the moir\'e zones compared to bare HOPG terraces as seen in Figure~\ref{fig:figur3}b and c. When the lineshapes of the STS curves are investigated, LDOS measurements of moir\'e areas have a larger curvature (more parabolic than linear) compared to the LDOS of HOPG terraces. This is an indicator of more metallic behaviour of the moir\'e zones compared to the HOPG terraces. Another important observation is the sharp van Hove like states in the calculated DOS of the moir\'e patterns, which were discussed intensively in the recent literature \cite{VanHove_HOPG_moire_Nature_Phy, Brihuega2012PRL, vHv_dirac_Chu2013, ARPES_no_VhV, CastroNeto_Bilayer_twist2007, Yan2014} . Several tight binding results were used to understand the nature of such features. According to previous reports energy values of the VHV singularities must follow a regular pattern as a function of the rotation angle of the top most layer. However, existence of a clear difference between GMPs and RMPs suggest a more complex nature and the answer to this complexity lies in the band structure of each moir\'e structure. 

In Figure~\ref{fig:figur3}d and e we compare the band structures of the same moir\'e pattern calculated using two different methods, ab initio planewave pseudopotential and tight binding calculations, respectively. These energy band structures are presented along the symmetry lines between the high symmetry points of the Brillouin zone (BZ) of the moir\'e superstructure unit cell of twisted bilayer graphene as indicated in Figure~\ref{fig:figur3}f. General features of moir\'e band structures can be summarized as follows: There are linearly dispersed double degenerate bands around the K-point of moir\'e BZ. However, the slope of these bands, which sets the Fermi velocity, decreases with decreasing twist angle as consistent with the previous studies \cite{Nanolett2010_magaud, magaud2}.  The interaction of the twisted top layer with its bottom layer does not open a band gap. Similarly, direct band gap at $\Gamma$ point also depends on the twist angle. Interestingly, there are almost linear and flat bands between the $\Gamma$ and M points only in the ab initio calculated band structures. Similarly, some critical bands, as we will discuss below, between M and M$'$ points as well as between K and $\Gamma$ points are split in ab initio calculations while they are degenerate in tight binding results. Since the number of atoms within the moir\'e superstructure unit cell depends on the twist angle and increases enormously with decreasing rotation angle, in the literature most of the studies were based on tight binding calculations for the rotations, which can be described by a few thousands atoms within the unit. However, as we noted above there are some critical differences between the ab initio and tight binding results. These may happen because of two reasons: (1) self consistent calculations are important for correct description of charge density distribution and (2) the inter-layer interaction, so the inter-layer hopping term needs to be adjusted for tight binding calculations. We must once more point out that with the rotation of the top layer the continuously changing observable in the super structure is the GMP not the RMP. As a result the band structure calculations must be investigated by keeping the atomistic structure of these patterns in mind. 

\subsection{Reason of van Hove singularities in the DOS of moir\'e patterns from their respective band structures}

We focused on one of the special cases in which RMP and GMP coincides and compared it to one of the more common cases where RMP is $\sqrt3$ times the GMP.  Different types of critical points exist in the dispersion of two-dimensional structures like graphene.  When the critical point is a minimum or a maximum, a step function singularity occurs in the density of states, while at the saddle points, a logarithmic singularity arises. Hence, in general, saddle points are responsible for sharp peaks in the DOS \cite{vanhove}. A nice example of this van Hove (VHV) singularity is the sharp peaks around +1.5 eV and -2 eV at DOS of graphene (or graphite), which are due to the saddle points at M point of its BZ.  In Figure~\ref{fig:figur4}a the band structure of $p_{1}$=1 and $q_{1}$=10 moir\'e is shown, which is one of the special cases where RMP is the same as GMP. In this system the rotation of the layers with respect to each other is 9.43$^{\circ}$ and the moir\'e period is 1.496 nm. When we inspect the band structure and corresponding DOS thoroughly, it is evident that the VHV originates from the saddle point of band edges between $\Gamma$ and M points, and in this case it is closer to the M point. We also show and discuss this in Figure~\ref{fig:figur4}d \footnote{The origin of the van Hove singularities in the moir\'e system is described further in the supplementary video}. In Figure~\ref{fig:figur4}b we present the band structure for one of the common cases, namely $p_{1}$=1 and $q_{1}$=12, which has a rotation angle of 7.93$^{\circ}$ with GMP of 1.78 nm and RMP of 3.082 nm. Again upon inspection of the band structure we see that the VHV of this case originate from the saddle point of the bands along the line connecting $\Gamma$ and K points. We can mark this between the M and M$'$ points as well (see Figure~\ref{fig:figur3}d and f). This is better visible in Figure~\ref{fig:figur4}e. 

From our results it is clear that the VHV singularities occurring in the DOS of different moir\'e patterns do not have the same origin in their band structure and this is clearly related to the existence of GMPs and RMPs. So far this has not been recognized in the reported studies. Rather, the emergences of VHV singularities were associated to intersection of Dirac cones of top and bottom layer exactly at M point of moir\'e BZ. In Figure~\ref{fig:figur4}c, we present the iso-surface of the $\pi$-band of the graphene band structure in the irreducible part of the graphene BZ around the K point. Essentially, this is a cosine band within tight binding approach and the linear part around the K point is known as Dirac cone. However, the Dirac cone diverges from a circular cone as the energy increases. We want to emphasize that the energies at which the VHV singularities occur are not close to the Fermi energy. On this band diagram, we highlighted the energies corresponding to the VHV singularities of the twisted bilayer graphene. The inner most curve corresponds to 100 meV. At this low energy the cross section is circular, which is in the linear part of the Dirac cone. Apart from these cases this constant energy cross section of $\pi$-band of graphene is more triangular, which becomes clear at higher energies. Apparently, these are outside the low excitation region where the continuum theories are applicable. If we had linear Dirac cones, then the intersection of these cones from top and bottom layer should occur exactly at M point of moir\'e BZ. As more of such non-circular cones intersect each other, the intersection generating the VHV singularities diverges form linearity (M point is a special case). The cone intersection does not occur as commonly represented for bilayer graphene at small energy values around the K point. Consequently the cones emanating from the K points do not intersect at the same places in the BZ for every rotation angle, as exemplified in Figure~\ref{fig:figur4}d and e. Furthermore, this intersection point depends on the orientation of the moir\'e BZ, so the origin of VHV singularity in the special cases where GMP=RMP and general cases where they are not equal are completely different. 

\section{Conclusions}

Our combined STM and ab initio studies on the moir\'e patterns on HOPG surfaces clearly showed the importance of proper atomistic modelling of the moir\'e patterns on such systems. Although DOS of RMPs and GMPs might appear similar, the band structures of these systems differed in detail. The emerging VHV singularities in their respective DOS had completely different origins from the energy bands of corresponding moir\'e systems. We have shown that the corrugations of different moir\'e patterns change as a function of the tunnel junction bias differently from each other in STM measurements. Although STM may not have the necessary spatial resolution to image the localized charge distribution of these moir\'e structures, we have evidence, both theoretically and experimentally, that the corrugation of the moir\'e patterns do not change monotonically. While we show that the corrugation of the moir\'e structures is not constant as a function of energy, we also show evidence for the non-graphene like behaviour of the top most layer of the moir\'e structures. Rather, we have shown that the graphene bilayer system generating the moir\'e structures behave as a large molecule.

In this study we have shown a very simple, reliable and repeatable method of producing moir\'e patterns on HOPG surfaces without damaging the crystals. We have clear evidence that the moir\'e/HOPG systems are perfect for being studied under ambient conditions with STM. Finally our results indicate that although moir\'e patterns on HOPG surfaces have mainly an electronic origin, the corrugation measurements show the hint of a minute structural origin, which is yet to be tested. 

Further measurements like dI/dV maps of the moir\'e patterns with high corrugation variance will also help understand the localized electronic structure of the moir\'e patterns. The corrugation variation of moir\'e systems are yet to be tested on SiC or graphene bilayers grown by CVD. Local electronic measurements on different moir\'es is another challenge for device applications. Our results have also direct relevance to moir\'e patterns observed on other layered systems such as hBN, or Graphene/hBN systems, especially in their modelling, which attracted recent attention \cite{fasolinoPRL} . Corrugation variation in such systems is another issue that awaits testing. The ease in the preparation of the samples we used for this study may help many others to pursue similar methodologies in similar layered systems. Our data has also direct relevance to graphene device literature where the identification of mono/bi-layer or multilayer graphene systems is invaluable while producing graphene based devices, which ideally function at room temperature and especially under ambient conditions, for real life applications.

\begin{acknowledgments}

This work was partially supported by T\"{U}B\.{I}TAK projects with grant numbers 109T687, 112T818,  113F005, and by ITU-BAP projects with numbers 33263 and 37705.

\end{acknowledgments}

\bibliography{moireRef_v16}

\end{document}